\documentstyle[12pt]{article}

\begin{document}

\hfill SOGANG-HEP 196/95

\hfill May 1995

\vspace{1cm}

\begin{center}
    {\large \bf Batalin-Tyutin Quantization of the
                Chiral Schwinger Model}
\end{center}

\vspace{1cm}

\begin{center}
    Jung-Ho Cha,
    Yong-Wan Kim\footnote{Electronic address: ywkim@physics.sogang.
    ac.kr}, Young-Jai Park\footnote{Electronic address:
    yjpark@physics.sogang.ac.kr},
    and Yongduk Kim,\\
    {\it Department of Physics and Basic Science Research Institute, \\
     Sogang University, C.P.O. Box 1142, Seoul 100-611, Korea}
\end{center}

\begin{center}
    Seung-Kook Kim \\
    {\it Department of Physics, Seo Nam University, Nam won 590-170, Korea}
\end{center}

\begin{center}
  Won T. Kim\footnote{Electronic address: wtkim@ewhahp3.ewha.ac.kr} \\
  {\it Department of Science Education and
  Basic Science Research Institute \\
   Ewha Women's University, Seoul 120-750, Korea}
\end{center}

\vspace{2cm}

\begin{center}
	{\bf ABSTRACT}
\end{center}

We quantize the chiral Schwinger Model by using the Batalin-Tyutin formalism.
We show that one can systematically construct the first class
constraints and the desired involutive Hamiltonian,
which naturally generates all secondary constraints.
For $a>1$, this Hamiltonian gives
the gauge invariant Lagrangian including the well-known Wess-Zumino terms,
while for $a=1$ the corresponding Lagrangian has the additional new type
of the Wess-Zumino terms,
which are irrelevant to the gauge symmetry.

\vspace{3cm}

\newpage

\begin{center}
\large{\bf I. Introduction}
\end{center}

Batalin and Fradkin (BF) [1] had proposed a new kind of
quantization procedure for second class constraint systems.
When combined with Batalin et al. (BFV) [2]
formalism for first class constraint systems,
the BFV formalism is particularly powerful for deriving
a covariantly gauge-fixed action in configuration space.
Fujiwara et al. (FIK) [3] have proposed an improved
treatment of anomalous gauge theories based on the BF formalism.
We have applied the FIK method to the bosonized
chiral Schwinger model (CSM) [4].
Recently, Banerjee, Rothe, and Rothe [5] have pointed out
that the FIK analyses [3,4] are not a systematic
application of the BFV formalism.
After their work,
Banerjee [6] has systematically applied Batalin-Tyutin (BT)
Hamiltonian method [7] to the second class constraint system of
the abelian Chern-Simons (CS) field theory [8-10].
As a result, he has obtained the new type of
an abelian Wess--Zumino (WZ) action,
which is irrelevant to the gauge symmetry.
Very recently, we have quantized the nonabelian case [11],
and the abelian self-dual massive theory [12] by using the BT formalism.
As shown in these works,
the nature of second class constraint algebra also originates
from the symplectic structure of the CS term
as well as the local gauge symmetry breaking effect.
There are some other interesting examples in this approach [13].

On the other hand, there has been a great progress
in the understanding of the physical meaning of
anomalies in quantum field theory through the study of the CSM.
Jackiw and Rajaraman [14]
showed that a consistent and unitary,
quantum field theory is even possible in the gauge non-invariant formulation.
Alternatively, a gauge invariant version [15] can be obtained by adding
a Wess-Zumino action to the gauge non-invariant original theory,
as was proposed by Faddeev and Shatashvili [16].
Since their works, the CSM have been still
analyzed by many authors as an archetype of anomalous gauge theory [4,5,17].

In the present paper, we shall apply the BT method to the bosonized CSM
having still novel features.
In Sec. II, we consider the bosonized
CSM with $a>1$, which has two second class constraints.
Through the BT analysis,
we will obtain the well-known WZ term to cancel the usual gauge anomaly
after we convert the original second class system into the fully
first class one.
In Sec. III, we consider the bosonized CSM for
$a=1$, which has four second class constraints.
In contrast to the $a>1$ case, we will obtain
an additional new WZ action,
which cannot be obtained in the usual path-integral framework,
as well as the usual WZ action needed to cancel the gauge anomaly.
In fact, the usual WZ action is not enough to make the second class
system the first class one for $a=1$.
Sec. IV is devoted to a conclusion.

\vspace{1cm}

\newpage

\begin{center}
\large{\bf II. CSM in the case of $a>1$}
\end{center}

In this section, we consider the bosonized CSM model in the case of $a>1$
[16]
\begin{equation}
    S_{CSM} ~=~ \int d^2x~\left[
			  -\frac{1}{4}F_{\mu \nu}F^{\mu \nu}
			  +\frac{1}{2}\partial_{\mu}\phi\partial^{\mu}\phi
			  +eA_{\nu}(\eta^{\mu \nu}
			  -\epsilon^{\mu \nu})\partial_{\mu}\phi
              +\frac{1}{2}ae^{2}A_{\mu}A^{\mu}~\right],
\end{equation}
where $\eta^{\mu\nu}=\mbox{diag.(1,-1)}$, $\epsilon^{01}=1$, and
$a$ is a regularization ambiguity [14], which is defined for calculating
the fermionic determinant of the fermionic CSM.
The canonical momenta are given by
\begin{eqnarray}
    \Pi^0~&=&~0, \nonumber \\
	\Pi^1~&=&~F_{01}~=~\dot{A}_1~-~\partial_{1}A_0, \nonumber \\
	\Pi_\phi~&=&~\dot{\phi}~+~e(A_0~-A_1),
\end{eqnarray}
where the overdot means the time derivative.
Following the usual Dirac's standard procedure [19],
there are one primary constraint
\begin{equation}
    \Omega_1 \equiv \Pi^0 \approx 0,
\end{equation}
and one secondary constraint
\begin{equation}
    \Omega_2 \equiv \partial_1 \Pi^1 + e\Pi_\phi + e\partial_1 \phi
                       + e^2 A_1 + (a-1)e^2 A_0 \approx 0.
\end{equation}
This constraint is obtained by conserving $\Omega_1$
with the total Hamiltonian
\begin{equation}
    H_T = H_c + \int dx~u\Omega_1 ,
\end{equation}
where $H_c$ is the canonical Hamiltonian as follows
\begin{eqnarray}
    H_c &=& \int\!dx~\left[~
			\frac{1}{2}(\Pi^{1})^2 + \frac{1}{2}(\Pi_{\phi})^2
             + \frac{1}{2}(\partial_{1}\phi)^2 - e(\Pi_{\phi}
             + \partial_{1}\phi)(A_0 - A_1) \right. \nonumber \\
        &&~~~~~~ \left. - A_{0}\partial_{1}\Pi^1
               - \frac{1}{2}ae^{2} \{ (A_0)^2 - (A_1)^2 \}
               + \frac{1}{2}e^{2}(A_0 - A_1)^2~\right],
\end{eqnarray}
and we denote a Lagrange multiplier $u$.
Note that by fixing the Lagrange multiplier $u$ as follows
\begin{equation}
	u~=~\partial_{1}A_1~-~\frac{1}{a-1}\Pi^1,
\end{equation}
no further constraints are generated via this procedure.
Then, the constraints $\Omega_\alpha (\alpha=1,2)$
form the second class algebra as follows
\begin{eqnarray}
	\Delta_{\alpha \beta}(x,y)
			&\equiv&
				\{ \Omega_{\alpha}(x), \Omega_{\beta}(y) \} \nonumber\\
            &=& e^2 (a-1)
				\left( \begin{array}{cc}
					  0   &    -1	         \\
					  1   &    0             \\
				\end{array} \right)
    \delta(x-y).
\end{eqnarray}

Following the BT approach [7],
we introduce new auxiliary fields $\Phi^\alpha$
in order to convert the second class constraint $\Omega_\alpha$ into
the first class in an extended phase space,
and assume that the Poisson algebra of the new fields is given by
\begin{equation}
   \{ \Phi^\alpha(x), \Phi^\beta(y) \} = \omega^{\alpha\beta}(x,y),
\end{equation}
where $\omega^{\alpha\beta}$ is an antisymmetric matrix.
Then, the modified constraints in the extended phase space are given by
\begin{equation}
  \tilde{\Omega}_\alpha(\Pi^\mu, A_\mu, \Phi^\alpha)
	 =  \Omega_\alpha + \sum_{n=1}^{\infty} \Omega_\alpha^{(n)};
			   ~~~~~~\Omega_\alpha^{(n)} \sim (\Phi^\alpha)^n
\end{equation}
satisfying the boundary condition,
$\tilde{\Omega}_\alpha(\Pi^\mu, A_\mu,0) = \Omega_\alpha$.
The first order correction term in the infinite series [7] is given by
\begin{equation}
  \Omega_\alpha^{(1)}(x) = \int dy ~X_{\alpha\beta}(x,y)\Phi^\beta(y),
\end{equation}
and the first class  constraint algebra
of $\tilde{\Omega}_\alpha$ requires the condition as follows
\begin{equation}
   \triangle_{\alpha\beta}(x,y) + \int dw~ dz
   ~X_{\alpha\mu}(x,w) \omega^{\mu\nu}(w,z) X_{\beta\nu}(z,y)= 0.
\end{equation}
As was emphasized in Ref. [6,11,12], there is a natural arbitrariness
in choosing $\omega^{\alpha\beta}$ and $X_{\alpha\beta}$
from Eq. (9) and Eq. (11), which corresponds to the canonical transformation
in the extended phase space [1,7].
Without any loss of generality, we take the simple solutions as
\begin{eqnarray}
	\omega^{\alpha\beta} (x,y) &=&
		\left( \begin{array}{cc}
            0         &    1          \\
           -1         &    0           \\
		\end{array} \right)
		\delta(x-y),      \nonumber \\
	X_{\alpha\beta} (x,y) &=&
		e \sqrt{a-1}
		\left( \begin{array}{cc}
			1         &    0           \\
			0         &    1           \\
		\end{array} \right)
		\delta(x-y),
\end{eqnarray}
which are compatible with Eq. (12),
and this choice considerably simplifies the algebraic manipulations.
As a result, using Eqs. (10), (11) and (13),
the new set of constraints is found to be
\begin{equation}
	\tilde{\Omega}_\alpha =
        \Omega_\alpha + e \sqrt{a-1} ~\Phi^\alpha,
\end{equation}
which are strongly involutive,
\begin{equation}
    \{ \tilde{\Omega}_\alpha, \tilde{\Omega}_\beta \} = 0.
\end{equation}
In other words, we can make the second class constraints the first
class by introducing the new fields in the extended phase space.
Therefore, we have all the first class constraints
in the extended phase space
by applying the BT formalism systematically.
Observe further that only $\Omega_\alpha^{(1)}$ contributes in the series
(10) defining the first class constraint.
All higher order terms given by Eq. (10) vanish
as a consequence of the proper choice (13).

Next, we derive the corresponding involutive Hamiltonian
in the extended phase space.
It is given by the infinite series [7],
\begin{equation}
    \tilde{H}(\Pi^\mu, A_\mu, \Phi^\alpha)
            = H_c + \sum_{n=1}^{\infty} H^{(n)};
    ~~~H^{(n)} \sim (\Phi^\alpha)^n
\end{equation}
satisfying the initial condition,
$\tilde{H}(\Pi^\mu, A_\mu, 0) = H_c$.
The general algebraic form for the involution of $\tilde{H}$ is given by
\begin{equation}
  H^{(n)} = -\frac{1}{n} \int dx dy dz~
		  \Phi^\alpha(x) \omega_{\alpha\beta}(x,y)
		  X^{\beta\gamma}(y,z) G_\gamma^{(n-1)}(z),
		  ~~~(n \geq 1),
\end{equation}
where the generating functionals $G_{\alpha}^{(n)}$ are given by
\begin{eqnarray}
  G_\alpha^{(0)} &=& \{ \Omega_\alpha^{(0)}, H_c \},  \nonumber  \\
  G_\alpha^{(n)} &=& \{ \Omega_\alpha^{(0)}, H^{(n)} \}_{\cal O}
			+ \{ \Omega_\alpha^{(1)}, H^{(n-1)} \}_{\cal O}
					   ~~~ (n \geq 1),
\end{eqnarray}
where the symbol ${\cal O}$ represents
that the Poisson brackets are calculated among the original variables,
{\it i.e.}, ${\cal O}=(\Pi^\mu, A_\mu)$.
Here, $\omega_{\alpha\beta}$ and $X^{\alpha\beta}$ are the inverse matrices
of $\omega^{\alpha\beta}$ and $X_{\alpha\beta}$, respectively.
Explicit calculations of $G^{(0)}_\alpha$ yield
\begin{eqnarray}
	G_1^{(0)} &=&  \Omega_2,   \\
    G_2^{(0)} &=&  e^2 \Pi^1 + e^2(a-1) \partial_1 A_1,
\end{eqnarray}
which are substituted in Eq. (17) to obtain $H^{(1)}$,
\begin{eqnarray}
	H^{(1)} &=& \frac{1}{e\sqrt{a-1}}\int dx [
				G_2^{(0)} \Phi^1 - \Omega_2 \Phi^2 ].
\end{eqnarray}
This is inserted back in Eq. (18) in order
to deduce $G_\alpha^{(1)}$ as follows
\begin{eqnarray}
	G_1^{(1)} &=& e\sqrt{a-1}~\Phi^2, \\
    G_2^{(1)} &=& -e\sqrt{a-1}~\partial_1^2 \Phi^1
			  + \frac{e^3}{\sqrt{a-1}}~ \Phi^1,
\end{eqnarray}
which then yield $H^{(2)}$ from Eq. (17),
\begin{equation}
    H^{(2)} =  \int dx
		 \left[
         \frac{1}{2}(\partial_1 \Phi^1 )^2
         + \frac{e^2}{2(a-1)}(\Phi^1)^2 - \frac{1}{2}(\Phi^2 )^2
		\right].
\end{equation}
Since $G_\alpha^{(n)} = 0 ~~(n \geq 2)$, the final expression for the
Hamiltonian after the $n=2$ finite truncations
is given by
\begin{equation}
	\tilde H = H_c + H^{(1)} + H^{(2)},
\end{equation}
which is strongly involutive with the first class constraints (14),
\begin{equation}
	\{\tilde{\Omega}_\alpha, \tilde H\} = 0.
\end{equation}

Before performing the momentum integrations to obtain the partition
function in the configuration space,
it seems to appropriate to comment on the strongly involutive
Hamiltonian (25).
If we use the above Hamiltonian (25), we cannot naturally generate
the first class Gauss' law constraint $\tilde{\Omega}_2$ from
the time evolution of the primary constraint $\tilde{\Omega}_1$.
Therefore, in order to avoid this problem as the case of the
self-dual massive model [12],
we use the equivalent first class Hamiltonian
without any loss of generality,
which only differs from the Hamiltonian (25)
by adding a term proportional to the first class constraint
$\tilde{\Omega}_2$ as follows
\begin{equation}
	\tilde{H}^{'} = \tilde{H}
			+ \int dx~\frac{1}{e\sqrt{a-1}} \Phi^2 \tilde{\Omega}_2.
\end{equation}
Then, this modified Hamiltonian $\tilde {H}^{'}$ consistently generates
the Gauss' law constraint
such that $\{ \tilde{\Omega}_1, \tilde{H}^{'} \} = \tilde{\Omega}_2$ and
$\{ \tilde{\Omega}_2, \tilde{H}^{'} \} = 0$.
Note that when we act this Hamiltonian on physical states,
the difference between $\tilde{H}'$ and $\tilde{H}$ is trivial
because such states are annihilated by the first class constraint.
Similarly, the equations of motion for observables ({\it i.e.},
gauge invariant variables) will also be unaffected by this difference
since $\tilde{\Omega}_2$ can be regarded as
the generator of the gauge transformations.

Now we derive the Lagrangian including the WZ term,
which describes the first class system,
corresponding to the Hamiltonian (27).
The first step is to identify the new variables
$\Phi^\alpha$ occurring in the extended phase space
as canonically conjugate pairs in the Hamiltonian formalism,
\begin{equation}
    \Phi^\alpha \equiv  \sqrt{a-1} \left( \theta,
                              \frac{1}{(a-1)} \Pi_{\theta} \right),
\end{equation}
satisfying Eqs. (9) and (13).
Then, the starting phase space partition function is given
by the Faddeev formula [20],
\begin{equation}
    Z =  \int  {\cal D} A_\mu {\cal D} \Pi^\mu
               {\cal D} \phi {\cal D} \Pi_\phi
               {\cal D} \theta {\cal D} \Pi_\theta
		  \prod_{\alpha,\beta = 1}^{2}
			\delta(\tilde{\Omega}_\alpha) \delta(\Gamma_\beta)
		  \det \mid \{\tilde{\Omega}_\alpha,\Gamma_\beta\} \mid
		  e^{iS'},
\end{equation}
where
\begin{equation}
	S'  =   \int d^2 x
		\left(
            \Pi^\mu {\dot A}_\mu
            + \Pi_\phi {\dot \phi}
            + \Pi_\theta {\dot \theta}
			- \tilde{\cal H'}
		\right)
\end{equation}
with the Hamiltonian density $\tilde{\cal H}'$ corresponding
to $\tilde H'$, which is now expressed in terms of
$\{\theta, \Pi_\theta\}$ instead of $\Phi^\alpha$. The gauge fixing
conditions $\Gamma_\alpha$ are chosen so that the determinant occurring
in the functional measure is nonvanishing.
Furthermore, $\Gamma_\alpha$
may be assumed to be independent of the momenta so that these are considered
as the Faddeev-Popov type gauge conditions.

Next, we perform the momentum integrations to obtain the
configuration space partition function.
First, the  $\Pi^0$ integration is trivially performed by exploiting
the delta function
$~~\delta(\tilde{\Omega}_1)~ =~ \delta[\Pi^0 + e(a-1)\theta]$ .
Then, after exponentiating the remaining delta function $
\delta(\tilde{\Omega}_2) =
    \delta[ \partial_1 \Pi^1 + e\Pi_\phi + e\partial_1 \phi
            + e^2 A_1 + (a-1)e^2 A_0 - e\Pi_\theta ] $ with
Fourier variable $\xi$ as $\delta(\tilde{\Omega}_2)=
\int{\cal D}\xi e^{-i\int d^2x\xi\tilde{\Omega}_2}$, and
transforming $A_0 \to A_0 + \xi$,
and integrating the other momentum variables $\Pi_\phi$ and $\Pi^1$,
we obtain the following intermediate action
\begin{eqnarray}
    S &=& \int d^2x~ \left[
		-\frac{1}{4}F_{\mu \nu}F^{\mu \nu}
		+\frac{1}{2}\partial_{\mu}\phi\partial^{\mu}\phi
		+eA_{\nu}(\eta^{\mu \nu}
		-\epsilon^{\mu \nu})\partial_{\mu}\phi
        +\frac{1}{2}ae^{2}A_{\mu}A^{\mu} \right. \nonumber \\
       &+& \theta
            \{ (a-1)e ( \partial_1 A_1 - \dot A_0 - \dot \xi )
              + \frac{1}{2}(a-1) \partial_1^2 \theta
              -e \epsilon^{\mu\nu} \partial_\mu A_\nu  \nonumber \} \\
       &+&    \left.
              \Pi_\theta \{ \dot\theta -e\xi-\frac{1}{2(a-1)}\Pi_\theta \}
              - \frac{1}{2} (a-1)e^2 \xi^2
              \right],
\end{eqnarray}
and the corresponding measure is given by
\begin{equation}
[{\cal D} \mu] = {\cal D} A_\mu
         {\cal D} \phi
		 {\cal D} \theta
         {\cal D} \Pi_\theta
		 {\cal D} \xi
		 \prod^2_{\beta = 1}
           \delta\left(\Gamma_{\beta}[A_0 + \xi, A_1, \theta]\right)
        \det \mid \{\tilde{\Omega}_{\alpha}, \Gamma_{\beta}\} \mid.
\end{equation}

At this stage, the original theory is simply reproduced,
if we choose the usual unitary gauge condition
\begin{equation}
    \Gamma_\alpha = ( \theta, \Pi_\theta).
\end{equation}
Note that this gauge fixing is consistent because when
we take the gauge fixing condition $\theta \approx 0$,
the condition $\Pi_\theta \approx 0$ is naturally generated from
the time evolution of $\theta$, {\it i.e.},
$\dot\theta = \{\theta, \tilde{H}' \}
= \frac{1}{(a-1)} \Pi_\theta \approx 0$.
Then, one can easily realize that
the new fields $\Phi^\alpha$ are nothing but the gauge degrees of
freedom, which can be removed by utilizing the gauge symmetry.

Finally, we perform the Gaussian integration over $\Pi_\theta$.
Then all terms including $\xi$ in the action are canceled out,
the resultant action is obtained as follows
\begin{eqnarray}
    S &=& S_{CSM} + S_{WZ} ~;           \nonumber \\
    S_{CSM} &=&  \int d^2x~ \left[
		-\frac{1}{4}F_{\mu \nu}F^{\mu \nu}
		+\frac{1}{2}\partial_{\mu}\phi\partial^{\mu}\phi
		+eA_{\nu}(\eta^{\mu \nu}
		-\epsilon^{\mu \nu})\partial_{\mu}\phi
        +\frac{1}{2}ae^{2}A_{\mu}A^{\mu}~\right], \nonumber \\
    S_{WZ} &=&  \int d^2x \left[
                \frac{1}{2}(a-1)\partial_\mu \theta \partial^\mu \theta
                - e \theta \{ (a-1)\eta^{\mu\nu} + \epsilon^{\mu\nu} \}
                    \partial_{\mu} A_\nu  \right],
\end{eqnarray}
where $S_{WZ}$ is the well-known WZ term, which is needed to cancel the
gauge anomaly.
On the other hand,
the corresponding Liouville measure just comprises the configuration
space variables as follows
\begin{equation}
[{\cal D} \mu] = {\cal D} A_\mu
         {\cal D} \phi
		 {\cal D} \theta
		 {\cal D} \xi
		 \prod^2_{\beta = 1}
           \delta\left(\Gamma_{\beta}[A_0 + \xi, A_1, \theta]\right)
        \det \mid \{\tilde{\Omega}_{\alpha}, \Gamma_{\beta}\} \mid.
\end{equation}
Starting from the Lagrangian (34),
we can easily reproduce the same set of all the first class constraints
(14) and the modified Hamiltonian (27)
effectively equivalent to the strongly involutive Hamiltonian (25).

Now, it seems appropriate to comment on the momentum integration by taking
the different order.
After integrating $\Pi^0$, exponentiating $\delta(\tilde{\Omega}_2)$,
transforming $A_0 \rightarrow A_0 + \xi$,
let us integrate $\Pi_\theta$ and $\Pi^1$ in order.
Then we obtain another intermediate action as follows
\begin{eqnarray}
S &=& \int d^2 x ~\left[ -\frac{1}{4}F_{\mu\nu}F^{\mu\nu}
                + \frac{1}{2}(a-1)\partial_\mu \theta \partial^\mu \theta
                - e \{ (a-1) \eta^{\mu\nu} + \epsilon^{\mu\nu} \}
                      \theta \partial_\mu A_\nu \right. \nonumber \\
      ~~~~~~~~~~&&   + \frac{1}{2} a e^2 A_\mu A^\mu
                 + \phi \{ \frac{1}{2} \partial_1^2 \phi
                      + e(\epsilon^{\mu\nu}-\eta^{\mu\nu})
                         \partial_\mu A_\nu \}
                                               \nonumber \\
      ~~~~~~~~~~&&  \left.
                       - \frac{1}{2} \{ \Pi_\phi - e (A_0-A_1) \}
                         \{ \Pi_\phi - e (A_0-A_1) - 2 \dot\phi \}
                    \right],
\end{eqnarray}
and the corresponding measure is given by
\begin{equation}
[{\cal D} \mu] = {\cal D} A_\mu
         {\cal D} \phi
         {\cal D} \Pi_\phi
         {\cal D} \theta
         {\cal D} \xi
		 \prod^2_{\beta = 1}
           \delta\left(\Gamma_{\beta}[A_0 + \xi, A_1, \phi, \theta]\right)
        \det \mid \{\tilde{\Omega}_{\alpha}, \Gamma_{\beta}\} \mid.
\end{equation}
In this case, if we choose the matter
gauge fixing condition $\phi \approx 0$ [18],
then we obtain the consistency condition from the time evolution of $\phi$
as follows
\begin{equation}
\dot\phi = \{ \phi,~ \tilde{H}' \} = \Pi_\phi - e (A_0-A_1) \approx 0.
\end{equation}
Therefore, if we choose the above unitary gauge as follows
\begin{equation}
\Gamma_\alpha = \left( \phi, \Pi_\phi - e A_0 + e A_1 \right),
\end{equation}
we easily reproduce the equivalent anomalous CSM [18].
Note that the WZ field $\theta$
becomes a dynamical field instead of the original matter
field $\phi$ in this equivalent model.
Furthermore, if we perform the Gaussian integration over $\Pi_\phi$
without choosing this gauge at this stage,
we obtain the same resultant action (34).
Therefore, we have explicitly shown that
the final desired result is independent of the order of
the momentum integration.
Furthermore, if we add a term proportional to the constraint
$ \tilde{\Omega}_1$,
which is trivial when acting on the physical
Hilbert space, to the Hamiltonian (27) as follows
\begin{equation}
\tilde{H}_c = \tilde{H}'+ \int dx ~u \tilde{\Omega}_1,
\end{equation}
one can exactly reproduce the BFV
Hamiltonian $\tilde{H}_c$ of the CSM obtained in Ref. [4]
with the exactly same first class constraints (14) with the new fields (28).
Then, one can easily reconstruct
the covariant effective action of the CSM, which is invariant under the BRST
transformation [4].

\begin{center}
\large{\bf III. CSM in the case $a=1$}
\end{center}

In this section, we consider the CSM in the case of $a=1$,
which is given by
\begin{equation}
   S ~=~ \int d^2 x ~\left[
		-\frac{1}{4}F_{\mu \nu}F^{\mu \nu}
        +\frac{1}{2}\partial_{\mu}\phi\partial^{\mu}\phi
        +eA_{\nu}(\eta^{\mu \nu}
        -\epsilon^{\mu \nu})\partial_{\mu}\phi
        +\frac{1}{2}e^{2}A_{\mu}A^{\mu}~\right].
\end{equation}
The canonical momenta are given by
\begin{eqnarray}
	\Pi^0~&\approx&~0, \nonumber \\
	\Pi^1~&=&~F_{01}~=~\dot{A}_1~-~\partial_{1}A_0, \nonumber \\
	\Pi_\phi~&=&~\dot{\phi}~+~e(A_0~- A_1).
\end{eqnarray}
There are one primary constraint
\begin{equation}
    \Omega_1 \equiv \Pi^0 \approx 0,
\end{equation}
and three secondary constraints
\begin{eqnarray}
	\Omega_2 &\equiv& \partial_1 \Pi^1 + e\Pi_\phi + e\partial_1 \phi
              + e^2 A_1 , \nonumber \\
    \Omega_3 &\equiv& e^2 \Pi^1 , \nonumber \\
    \omega_4 &\equiv& -e^3 \Pi_\phi -e^3 \partial_1 \phi
              +e^4 A_0 - 2e^4 A_1 .
\end{eqnarray}
Note that these constraints are obtained by conserving
the constraints with the total Hamiltonian,
\begin{equation}
    H_T = H_c + \int dx~ u \Omega_1 ,
\end{equation}
where $H_c$ is the canonical Hamiltonian,
\begin{eqnarray}
    H_c &=& \int d x \left[
        \frac{1}{2}(\Pi^1)^{2}+\frac{1}{2}(\Pi_{\phi})^2
        +\frac{1}{2}(\partial_{1}\phi)^2-e(\Pi_{\phi}
        +\partial_{1}\phi)(A_{0}~-A_{1}) \right. \nonumber \\
       ~~~~~~~~~~~~~~~~ && \left.
           A_{0}\partial_{1}\Pi^{1}-e^{2}A_{0}A_{1}
              +e^{2}(A_{1})^2~\right],
\end{eqnarray}
and we denote the Lagrange multiplier $u$. By fixing the Lagrange
multiplier $u$ as follows
\begin{equation}
	u~= ~\frac{1}{e}\partial_{1}\Pi_{\phi}
        ~+~\frac{1}{e} \partial_1^2 \phi
		~+~ 2 \Pi^{1}~+~2\partial_{1}A_{1},
\end{equation}
no further constraints are generated via this procedure.
We find that all the constraints are fully second class constraints.
However, in order to carry out the simple algebraic manipulations,
it is essential to redefine $\omega_4$ by using $\Omega_1$ as
follows
\begin{eqnarray}
	\Omega_4 &\equiv&
				\omega_4 + e^2 \partial_1 \Omega_1 \nonumber\\
			 &=& -e^3 \Pi_\phi - e^3\partial_1 \phi
				 +e^4 A_0 - 2e^4 A_1 + e^2\partial_1 \Pi^0 ,
\end{eqnarray}
although the redefined constraints are still completely the second class in
contrast to the CS theories [6,11].
Otherwise, one will have a complicated constraint algebra including the
derivative terms which are difficult to handle.
Then, the simplified second class constraint algebra
for $\Omega_\alpha (\alpha=1,\cdot\cdot\cdot,4)$ is given by
\begin{eqnarray}
    \Delta_{\alpha\beta}(x,y)
		&\equiv&  \{ \Omega_{\alpha}(x), \Omega_{\beta}(y) \} \nonumber\\
		&=& e^4  \left( \begin{array}{cccc}
					  0   &  0   &  0    &  -1     \\
					  0   &  0   &  1    &  0          \\
					  0   & -1   &  0    & 2e^2	\\
					  1 &  0     & -2e^2 &  0
				 \end{array} \right)
  \delta(x-y).
\end{eqnarray}

Following the BT approach [7], we introduce the matrix (9),
which is compatible with the new fields $\Phi^\alpha$ as follows
\begin{equation}
	\omega^{\alpha\beta} (x,y) =
		\left( \begin{array}{cccc}
			0 	&  0  &  0  &  1    \\
			0 	&  0  &  1  &  0    \\
			0 	& -1  &  0  &  0    \\
		   -1 	&  0  &  0  &  0
	   \end{array}
   \right)
  \delta(x-y).
\end{equation}
Then the other matrix $X_{\alpha\beta}$ in Eq. (11) is easily obtained by
solving Eq. (12) with $\Delta_{\alpha\beta}$ given by Eq. (49),
\begin{equation}
	X_{\alpha\beta} (x,y) =
   e^2		\left( \begin{array}{cccc}
			-1  	&    0     &    0   &    0    \\
			0       &  -1      &    0   &    0    \\
			e^2     &    0     &    1   &    0    \\
			0       &   e^2    &    0   &   -1
		\end{array} \right)
	\delta(x-y).
\end{equation}
Similar to the $a>1$ case, there is also an arbitrariness in choosing
$\omega^{\alpha\beta}$,
which would naturally be manifested in Eq. (50).
However, as has also been evidenced in other calculations [6,11,14],
these choices of Eqs. (50) and (51) give
remarkable algebraic simplifications for the case of $a=1$.
Using Eqs. (10), (11), (50) and (51),
the new set of constraints is found to be
\begin{eqnarray}
	\tilde{\Omega}_1 &=& \Omega_1 ~-~e^2 \Phi^1~,  \nonumber\\
	\tilde{\Omega}_2 &=& \Omega_2 ~-~e^2 \Phi^2~, \nonumber \\
	\tilde{\Omega}_3 &=& \Omega_3 ~+~e^4 \Phi^1 ~+~ e^2 \Phi^3~,\nonumber \\
	\tilde{\Omega}_4 &=& \Omega_4 ~+~e^4 \Phi^2 ~-~ e^2 \Phi^4 ~,
\end{eqnarray}
which are strongly involutive as those should be
\begin{equation}
\{ \tilde{\Omega}_{\alpha}, \tilde{\Omega}_{\beta} \} = 0.
\end{equation}
Recall the $\Phi^{\alpha}$ are the new variables satisfying the algebra
(9) with $\omega^{\alpha\beta}$  given  by Eq. (50).
Therefore, we obtain the fully first class constraint system
in the extended phase space.

The next step is to obtain the involutive Hamiltonian including
the new fields $\Phi^\alpha$.
It is noteworthy that there are only two terms $\Omega_\alpha$ and
$\Omega_\alpha^{(1)}$ in the expansion (52)
due to the intuitive choices (50) and (51).
The generating functionals $G_\alpha^{(n)}$
are obtained from Eq. (18) as follows,
\begin{eqnarray}
	G_i^{(0)} &=& \Omega_{i+1}  ~~~~~~~(i=1,2)~, \nonumber \\
	G_3^{(0)} &=& \Omega_4 ~-~ e^2 \partial_1 \Omega_1~, \nonumber \\
    G_4^{(0)} &=& e^2 \partial_1^2 \Pi^1
			  ~-~2e^4 \Pi^1 ~-~ e^4 \partial_1 A_1 ,
\end{eqnarray}
which are substituted in Eq. (17) to obtain $H^{(1)}$,
\begin{equation}
    H^{(1)} = - \frac{1}{e^2} \int dx~ [
            \Phi^1 ( e^2 \Omega_3 + G_4^{(0)} )
           -\Phi^2 (e^2 \Omega_2 +\Omega_4 - e^2 \partial_1 \Omega_1 )
           - \Phi^3 \Omega_3
           - \Phi^4 \Omega_2
		 ].
\end{equation}
This is inserted back in Eq. (18) to deduce $G_\alpha^{(1)}$ as follows
\begin{eqnarray}
	G_1^{(1)} &=& -e^2 \Phi^2, \nonumber \\
	G_2^{(1)} &=& e^4 \Phi^1 + e^2 \Phi^3, \nonumber\\
	G_3^{(1)} &=& e^4 \partial_1 \Phi^1 + e^4 \Phi^2 - e^2 \Phi^4,
					  \nonumber\\
	G_4^{(1)} &=& -2e^6 \Phi^1 + 2e^4 (\partial_1 )^2 \Phi^1
				  +e^4 \partial_1 \Phi^2 - 2e^4 \Phi^3,
\end{eqnarray}
which then yield $H^{(2)}$
\begin{equation}
	H^{(2)} = \int dx
		\left[
			\frac{1}{2}e^4 (\Phi^1 )^2 +e^2 \Phi^1 \Phi^3
			+e^2(\partial_1 \Phi^1 )^2 -e^2 \Phi^1 \partial_1 \Phi^2
			-\Phi^2 \Phi^4 + \frac{1}{2}(\Phi^3 )^2
	   \right].
\end{equation}
Since $G_\alpha^{(n)} = 0 ~~(n \geq 2)$,
after the $n=2$ finite truncations
the final expression for the desired Hamiltonian is given by
\begin{equation}
	\tilde H = H_c + H^{(1)} + H^{(2)},
\end{equation}
which is involutive,
\begin{equation}
	\{\tilde{\Omega}_\alpha, \tilde H\} = 0.
\end{equation}
According to the usual BT formalism,
this formally completes the operatorial, abelian conversion of the original
second class system with the Hamiltonian $H_c$ and the constraints
$\Omega_\alpha$ into the first class
with the Hamiltonian $\tilde H$ and the constraints $\tilde{\Omega}_\alpha$.

However, similar to the $a>1$ case,
if we use the above Hamiltonian, we cannot naturally generate
the first class constraints $\tilde{\Omega}_i (i=2,3,4)$ from
the time evolution of the primary constraint $\tilde{\Omega}_1$.
In order to avoid this situation,
we also use another equivalent first class Hamiltonian
without any loss of generality,
which differs from the involutive Hamiltonian (58)
by adding terms proportional to the first class constraint
$\tilde{\Omega}_\alpha$ as follows
\begin{equation}
	\tilde{H}^{'} = \tilde{H}
                    + \Phi^1 \tilde{\Omega}_3
                    - \frac{1}{e^2} \Phi^2 (\tilde{\Omega}_3
                        + e^2 \tilde{\Omega}_2 )
                    - \frac{1}{e^2} \Phi^3 \tilde{\Omega}_3
                    - \frac{1}{e^2} \Phi^4 \tilde{\Omega}_2,
\end{equation}
which is easily found through the simple algebraic manipulations of
the new fields $\Phi^\alpha$ with $\omega^{\alpha\beta}$.
Then, this Hamiltonian $\tilde {H}^{'}$ automatically generates
the first class constraints
such that $\{ \tilde{\Omega}_i, \tilde{H}^{'} \} =
\tilde{\Omega}_{i+1}$ $(i=1,2,3)$ and
$\{ \tilde{\Omega}_4, \tilde{H}^{'} \} =0.$

We now extract out the Lagrangian
corresponding to the Hamiltonian (60).
The first step is to identify the new variables
$\Phi^\alpha$ as canonically conjugate pairs in the Hamiltonian formalism,
\begin{equation}
    \Phi^\alpha \equiv ( \frac{1}{e} \theta,
                  -\frac{1}{e} \rho,
                  -e\Pi_\rho,
                   e\Pi_\theta )
\end{equation}
satisfying Eqs. (9) and (50).
The starting phase space partition function is then given
by the Faddeev formula [20],
\begin{equation}
    Z =  \int  {\cal D} A_\mu  {\cal D} \Pi^\mu
               {\cal D} \phi {\cal D} \Pi_\phi
               {\cal D} \theta {\cal D} \Pi_\theta
			   {\cal D} \rho   {\cal D} \Pi_\rho
		 \prod_{\alpha,\beta = 1}^{4}
				\delta(\tilde{\Omega}_\alpha)\delta(\Gamma_\beta)
		 \det \mid \{\tilde{\Omega}_\alpha,\Gamma_\beta\} \mid
		 e^{iS'},
\end{equation}
where
\begin{equation}
	S'  =  \int d^2x \left(
                \Pi^\mu {\dot A}_\mu
                + \Pi_\phi {\dot \phi}
                + \Pi_\theta {\dot \theta}
				+ \Pi_\rho {\dot \rho} - \tilde{\cal H'}
		   \right)
\end{equation}
with the Hamiltonian density $\tilde{\cal H}'$ corresponding
to $\tilde H'$, which is now expressed in terms of
$\{\rho, \Pi_{\rho}, \theta, \Pi_\theta\}$
instead of $\Phi^\alpha$.
As in the previous section, the gauge fixing
conditions $\Gamma_\alpha$ may be also assumed to be independent
of the momenta so that these are considered as the Faddeev-Popov type
gauge conditions.

Next, to obtain the partition function in the configuration space,
we perform the momentum integrations by taking the proper order for
more simpler calculation without any loss of generality.
First, the  $\Pi^0$, $\Pi_\theta$ and $\Pi_\rho $ integrations are
trivially performed by exploiting
the delta functions
$~~\delta(\tilde{\Omega}_1)~ =~\delta[\Pi^0 -e\theta]$,
$~~\delta(\tilde{\Omega}_4)~ =~
	\delta[-e^3 \Pi_\phi - e^3 \partial_1 \phi + e^4 A_0
			-2e^4 A_1 -e^3 \rho -e^3 \Pi_\theta + e^3 \partial_1 \theta	]$,
and $~~\delta(\tilde{\Omega}_3)~ =~
	\delta[e^2 \Pi^1 +e^3 \theta - e^3 \Pi_\rho] $.
Then, after exponentiating the remaining delta function $
\delta(\tilde{\Omega}_2) =
	\delta[ \partial_1 \Pi^1 + e\Pi_\phi + e\partial_1 \phi
			+ e^2 A_1 + (a-1)e^2 A_0 - e\Pi_\theta ] $ with
Fourier variable $\xi$ as $\delta(\tilde{\Omega}_2)=
\int{\cal D}\xi e^{-i\int d^2x\xi\tilde{\Omega}_2}$ and
transforming $A_0 \to A_0 + \xi$, we obtain the action as follows
\begin{eqnarray}
S &=& \int d^2x  \left[
               - \frac{1}{2} \partial_\mu \phi \partial^\mu \phi
               + \frac{1}{2} e^2 A_\mu A^\mu
               + e(\eta^{\mu\nu}-\epsilon^{\mu\nu}) A_\nu \partial_\mu \phi
                 \right.          \nonumber \\
     ~~~~~~&&  + \theta \partial_1^2 \theta
               + \rho \partial_1 \theta - \dot{\theta} \partial_1 \phi
               - e A_1 \rho - \frac{1}{2} \rho^2
               - \rho \partial_1 \phi - e A_1 \dot{\theta}
               + \dot{\theta} \partial_1 \theta
               + \dot{\rho} \theta
                \nonumber \\
     ~~~~~~&& \left.
               + e \epsilon^{\mu\nu} \theta \partial_\mu A_\nu
               + \frac{1}{2} \dot{\theta}^2
               - \dot{\phi} \dot{\theta}
               - \rho \dot{\phi}
               + \Pi^1 (\dot{A}_1 - \partial_1 A_0
               + \frac{1}{e} \dot{\rho}
               + \frac{1}{e} \partial_1^2 \theta) \right],
\end{eqnarray}
and the corresponding measure is given by
\begin{equation}
[{\cal D} \mu] = {\cal D} A_\mu
         {\cal D} \phi
		 {\cal D} \theta
         {\cal D} \rho
         {\cal D} \Pi^1
		 {\cal D} \xi
         \prod^4_{\beta = 1}
    \delta\left(\Gamma_{\beta}[A_0 + \xi, A_1, \phi, \theta, \rho]\right)
        \det \mid \{\tilde{\Omega}_{\alpha}, \Gamma_{\beta}\} \mid.
\end{equation}

Finally, we perform the Gaussian integration over $\Pi^1$.
The resultant action is obtained as follows
\begin{eqnarray}
    S_{tot} &=& S_{CSM} + S_{WZ} + S_{NWZ} ~;      \nonumber \\
    S_{CSM} &=&  \int d^2x~ \left[
		-\frac{1}{4}F_{\mu \nu}F^{\mu \nu}
		+\frac{1}{2}\partial_{\mu}\phi\partial^{\mu}\phi
		+eA_{\nu}(\eta^{\mu \nu}
		-\epsilon^{\mu \nu})\partial_{\mu}\phi
        +\frac{1}{2}e^{2}A_{\mu}A^{\mu}~\right], \nonumber \\
    S_{WZ} &=&  - \int d^2x  [ e \theta \epsilon^{\mu\nu}
                          \partial_\mu A_\nu ], \nonumber \\
    S_{NWZ} &=&  \int d^2x
            \left[ - (\partial_1 \phi + eA_1 + \dot\phi -\partial_1 \theta)
                   (\dot\theta+\rho)
                + \frac{1}{2} \{ (\dot\theta)^2 - \rho^2 \}
                     \right. \nonumber \\
         ~~~~&& \left.
           - \frac{1}{2e^2}(\dot\rho + \partial_1^2 \theta )^2
                    + e \theta \partial_\mu A^\mu
                \right],
\end{eqnarray}
where $S_{WZ}$ is a usual WZ term needed to cancel the gauge anomaly,
and $S_{NWZ}$ is a new type of WZ term, which is irrelevant to the
gauge symmetry.
Note that the new type of WZ term $S_{NWZ}$
as well as the well--known WZ term $S_{WZ}$ should be
needed to make the second class system into the first class.
On the other hand,
the corresponding nontrivial Liouville measure
just comprises the configuration space variables as follows
\begin{eqnarray}
[{\cal D} \mu] &=& {\cal D} A_\mu
         {\cal D} \phi
		 {\cal D} \theta
         {\cal D} \rho
         {\cal D} \xi
 \delta [F_{01} + \frac{1}{e}(\dot\rho + \partial_1^2 \theta) ] \nonumber \\
        &&~~ \prod^4_{\beta = 1}
           \{ \delta(\Gamma_{\beta}[A_0 + \xi, A_1, \phi, \theta, \rho]) \}
        \det \mid \{\tilde{\Omega}_{\alpha}, \Gamma_{\beta}\} \mid,
\end{eqnarray}
where $\delta [ F_{01} + e^{-1} ( {\dot \rho} + \partial^2_1 \theta) ]$
is expressed by $\int {\cal D} \Pi^1 e^{-i \int d^2 \! x [ F_{01} + e^{-1}
({\dot \rho} + \partial^2_1 \theta)] \Pi^1 }$.
Note that although the Maxwell term is disappeared
as the cases of the conventional phase space approach [2,21]
through the momentum integration due to the constraint $\Omega_3$
in Eq. (44), we reintroduce this term into the action (64)
because we have the $\delta$-function related
to $F_{01}$ in the measure part.
In contrast to the case of $a>1$, from the action $S_{tot}$
including the term of the phase space variable $\Pi^1$
due to the appearance of the $\delta$-function in the measure instead of the
Lagrangian (64) including only configuration space variables,
{\it i.e.}, at the level of action (66),
we can only reproduce the same set of all the first class constraints
(52) and the modified Hamiltonian (58).
However, although we have succeeded to obtain the first class system
for the case of $a=1$,
the final theory has not the gauge symmetry due to the presence of the
non-trivial $\delta$--function in the measure part.
Therefore, through this analysis we have learned that
in general the first class system do not always
need to have the gauge symmetry.
However, similar to the $a>1$ case,
if we add terms proportional to the constraints
$ \tilde{\Omega}_i, i.e.,$
\begin{equation}
\tilde{H}_c = \tilde{H}'+ \int dx~ [
    (u-\frac{1}{e^2}\partial_1\Omega_2 ) \tilde{\Omega}_1
    -\frac{1}{e^2}\partial_1 \Omega_1 \tilde{\Omega}_2 ],
\end{equation}
which is trivial when acting on the physical
Hilbert space, to the Hamiltonian (58),
we can exactly reproduce the BFV
Hamiltonian $\tilde{H}_c$ and the corresponding BRST invariant Lagrangian
of the CSM obtained in Ref. [4].

\begin{center}
\large{\bf IV. Conclusion}
\end{center}

We have quantized the bosonized CSM
having the different algebra of the constraints
depending on the regularization parameter $a$ by using the BT formalism.
We have shown that one can systematically construct the first class
constraints and the desired involutive Hamiltonian,
which naturally generates
all the secondary constraints including the Gauss constraint.
For $a>1$, this Hamiltonian gives
the gauge invariant Lagrangian including the well-known WZ terms,
while for $a=1$ the corresponding Lagrangian has the new type
of the WZ terms,
which are irrelevant to the gauge symmetry and cannot be obtained
in the usual path-integral framework,
as well as the usual WZ term to cancel the gauge anomaly.

\begin{center}
\section*{Acknowledgements}
\end{center}

The present study was supported in part by
the Sogang University Research Grants in 1995, and
the Basic Science Research Institute Program,
Ministry of Education, Project No. {\bf BSRI}-95-2414.

\newpage

\section*{References}

\begin{description}{}

\item{1.} I. A. Batalin, E. S. Fradkin:
           Phys. Lett. B180 (1986) 157; Nucl. Phys. B279 (1987) 514
\item{2.} E. S. Fradkin, G. A. Vilkovisky: Phys. Lett. B55
           (1975) 224;
           I. A. Batalin, G. A. Vilkovisky: Phys. Lett. B69
           (1977) 309
\item{3.} T. Fujiwara, Y. Igarashi, J. Kubo:
           Nucl. Phys. B341 (1990) 695
\item{4.} Y.-W. Kim et al.: Phys. Rev. D46 (1992) 4574
\item{5.} R. Banerjee, H. J. Rothe, K. D. Rothe:
           Phys. Rev. D49 (1994) 5438
\item{6.} R. Banerjee: Phys. Rev. D48 (1993) R5467
\item{7.} I. A. Batalin, I. V. Tyutin:
           Int. J. Mod. Phys. A6 (1991) 3255
\item{8.} Edited by S. Treiman et el.:
          Topological Investigations of Quantized Gauge Theories,
          Singapore: World Scientific 1985
\item{9.} G. Semenoff: Phys. Rev. Lett. 61 (1988) 517;
           G. Semenoff, P. Sodano: Nucl. Phys. B328 (1989) 753
\item{10.} R. Banerjee: Phys. Rev. Lett. 69 (1992) 17;
           Phys. Rev. D48 (1993) 2905
\item{11.} W. T. Kim, Y. -J. Park: Phys. Lett. B336 (1994) 376
\item{12.} Y.-W. Kim et al.:
           Phys. Rev. D51 (1995) 2943;
           E.-B. Park et al.: Mod. Phys. Lett. A10 (1995) to appear
\item{13.} N. Banerjee, S. Ghosh, R. Banerjee:
           Nucl. Phys. B417 (1994) 257; Phys. Rev. D49 (1994) 1996;
           R. Banerjee, H. J. Rothe, K. D. Rothe: Nucl. Phys. B426
           (1994) 129
\item{14.} R. Jackiw, R. Rajaraman: Phys. Rev. Lett. 54
           (1985) 1219; Phys. Rev. Lett. 54 (1985) 2060(E);
           N. K. Falck, G. Kramer: Ann. Phys. 176 (1987) 330;
           Z. Phys. C37 (1988) 321;
           K. Shizuya: Phys. Lett. B213 (1988) 298;
           K. Harada: Phys. Rev. Lett. 64 (1990) 139; Phys. Rev.
            D42 (1990) 4170
\item{15.} F. Schaposnik, C. Viallet:
           Phys. Lett. B177 (1986) 385;
           K. Harada, I. Tsutsui: Phys. Lett. B183 (1987) 311
\item{16.} L. D. Faddeev, S. S. Shatashvili: Phys. Lett. B167
           (1986) 225
\item{17.}  R. Rajaraman: Phys. Lett. B154 (1985) 305;
            H. O. Girotti, H. J. Rothe, K. D. Rothe:
            Phys. Rev. D34 (1986) 592;
            H. J. Rothe, K. D. Rothe: Phys. Rev. D40 (1989) 545;
            J. Sladkowski: Phys. Lett. B296 (1992) 361;
            J. -G. Zhou, Y. -G. Miao, Y. -Y. Liu:
            Mod. Phys. Lett. A9 (1994) 1273
\item{18.}  W. T. Kim et al.:
            Mod. Phys. Lett. A7 (1992) 411
\item{19.} P. A. M. Dirac: Lectures on quantum mechanics, New York:
           Yeshiba University Press 1964
\item{20.} L. D. Faddeev, V. N. Popov: Phys. Lett. B25 (1967) 29
\item{21.} I. Batalin, E. S. Fradkin: Phys. Lett. B128 (1983) 303
\end{description}
\end{document}